\documentclass[aps,prl,twocolumn,showpacs]{revtex4}

\usepackage{amssymb}
\usepackage{amsmath}
\usepackage{amsbsy}
\usepackage{bm}

\usepackage{graphicx}

\newcommand{\I}{\mathrm{i}}
\newcommand{\E}{\mathrm{e}}

\usepackage{color}
\DeclareMathOperator{\tr}{tr}
\begin{document}

\title{Pump-probe scheme for electron-photon dynamics in hybrid conductor-cavity systems}
\author{T. L. van den Berg}
\author{C. Bergenfeldt}
\author{P. Samuelsson}
\affiliation{Physics department, Lund University, Box 118, S-221 00 Lund, Sweden}
\date{\today}
\begin{abstract}
  Recent experiments on nanoscale conductors coupled to microwave
  cavities put in prospect transport investigations of electron-photon
  interplay in the deep quantum regime. Here we propose a pump-probe
  scheme to investigate the transient dynamics of individual
  electron-photon excitations in a double quantum dot-cavity
  system. Excitations pumped into the system decay via charge
  tunneling at the double dot, probed in real time. We investigate
  theoretically the short-time charge transfer statistics at the dot,
  for periodic pumping, and show that this gives access to vacuum Rabi
  oscillations as well as excitation dynamics in the presence of
  double dot dephasing and relaxation.
\end{abstract}
\pacs{73.23.Hk, 72.10.Di, 85.25.-j}
\maketitle

\textit{Introduction.--} Hybrid quantum electrodynamic (QED)
structures that incorporate nanoscopic conductors in superconducting
circuits have attracted a lot of interest over the last few
years. Several experiments have demonstrated coupling between both
single \cite{Frey11,Delbecq11,Delbecq13} and double
\cite{Frey112,Toida13,Petersson12,Viennot13,Deng13,Petta14} quantum
dots (DQDs) and a superconducting microwave cavity. Hybrid
conductor-cavity systems have large potential for applications in
quantum information technology \cite{Trif08,Cottet10,Jin12} and can be
used as a characterization tool for nanoscale systems
\cite{Cottet102,Basset13,Schiro13}. Moreover, they put in prospect
transport investigations as well as applications of electron-photon
interactions in the deep quantum limit
\cite{Lambert08,Bergenfeldt13,Lambert13,Contreras13,Xu13}. This
includes DQD masers \cite{Childress04,Jin11,Jin11}, entanglement
detection in Cooper-pair splitters \cite{Cottet12}, testbeds for
Franck-Condon physics \cite{Bergenfeldt12}, detection schemes for
Majorana fermions \cite{Schmidt13,Cottet13,Muller13}, photon emission
statistics \cite{Xu132} and efficient heat engines
\cite{Bergenfeldt132}.

A key feature of nanoscopic conductor-cavity system which to date has
received little attention is the dynamics of electron-photon
excitations. Besides being of fundamental interest and of importance
for conductor-cavity applications, knowledge of the dynamics is
instrumental in identifying and characterizing parasitic effects
responsible for excitation dephasing and relaxation
\cite{Wang09}. However, an experiment investigating the dynamics would
arguably require, in the same conductor-cavity system, both short time
excitation manipulation and detection. Importantly, in circuit QED
systems microwave photons can be generated and controlled in real time
with large accuracy \cite{hofheinz08,hofheinz09,Eichler11}. Moreover,
time resolved counting of electrons in quantum dot systems has been
demonstrated experimentally \cite{Gustavsson06,Flindt09}.

\begin{figure}[!h]
\includegraphics[width=1.0\linewidth]{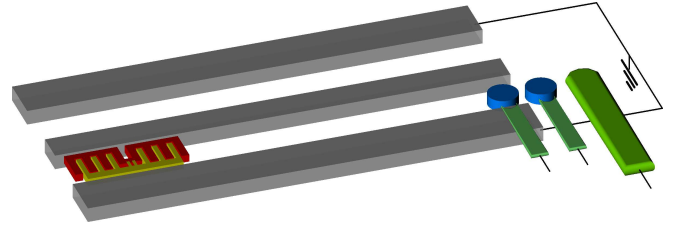}
\caption{(color online) Schematic of the hybrid system. An externally excited
  qubit (left) and a double quantum dot (right) are coupled to a coplanar
  superconducting microwave cavity. Further, one of the quantum dots is
  coupled to a lead.}
\label{fig_system}
\end{figure}

In this letter we combine these two capacities and propose a
\textit{pump-probe} scheme for investigating the transient dynamics of
electron-photon excitations in a hybrid conductor-cavity system (See
Fig. \ref{fig_system}). Single excitations are pumped into the system
via an externally driven qubit coupled to the cavity. The excitations
relax by electron tunneling at the DQD, probed by monitoring the dot
occupation in real time. We analyze theoretically the short time
charge transport statistics at the DQD for periodic pumping, and show
that it provides information on vacuum Rabi oscillations, as well as
excitation relaxation time scales. Moreover, the transport statistics
clearly displays how DQD dephasing and relaxation alter the dynamics,
key knowledge in the experimental efforts to reach the strong
conductor-cavity coupling regime.

\textit{System.--} The system under investigation consists of a DQD
coupled to a superconducting microwave cavity, and tunnel coupled to a
single lead electrode. The DQD, occupied with at most one excess
electron, contains two active levels. Moreover, an externally
controlled superconducting qubit, in Fig.~\ref{fig_system} exemplified
by a transmon \cite{transmon}, is coupled to the cavity. The DQD and
the qubit are coupled to the cavity with the same strength \(g_0\),
and kept at resonance with the fundamental cavity mode, at frequency
\(\omega_0\). In addition the characteristic impedance \(Z_0\) of the
cavity is taken much smaller than the resistance quantum
\(R_Q=h/e^2\). Under these conditions the DQD-cavity-qubit system is
described by the generalized Tavis-Cummings (TC) Hamiltonian
\begin{align}
\hat{H}_{S} &=  \hbar\omega_0 \hat{a}^{\dagger}\hat{a}+\frac{\hbar \omega_0}{2} \left(\hat{\sigma}_{z}+\hat{d}^\dagger_{e}\hat{d}_{e}-\hat{d}^{\dagger}_{g}\hat{d}_{g}\right) \nonumber \\
& + \hbar g_0
\left[\hat{a}^\dagger(\hat{\sigma}_{-}+\hat{d}^{\dagger}_{g}\hat{d}_{e}) +\hat
  {a}(\hat{\sigma}_{+}+\hat{d}^{\dagger}_{e}\hat{d}_{g}) \right]\,. \label{eq_H}
\end{align}
Here \(\hat a^\dag\) (\(a\)) is the photon creation (annihilation)
operator, and \(\hat{d}^{\dag}_{e/g}\) (\(\hat{d}_{e/g}\)) the
electron creation (annihilation) operators for the excited/ground
(bonding/anti-bonding) states of the DQD. The qubit operators
\(\hat{\sigma}_\pm = \hat{\sigma}_x\pm \I \hat{\sigma}_y\) and
\(\hat{\sigma}_z\) are expressed in Pauli matrices.  In the local
basis $|\pm\rangle$ denote the excited/ground state of the qubit,
$|e/g/0\rangle$ describe the DQD in the excited/ground state or
unoccupied and $|p\rangle$ the cavity state with $p$
photons. Throughout the paper we neglect qubit decoherence and cavity
losses, a reasonable assumption based on circuit QED experiments, see
e.g. \cite{Walraff,Schreier,Wang2}. A discussion of DQD dephasing and
relaxation is deferred to the last paragraph.
\begin{figure}[t!]
\includegraphics[width=\linewidth]{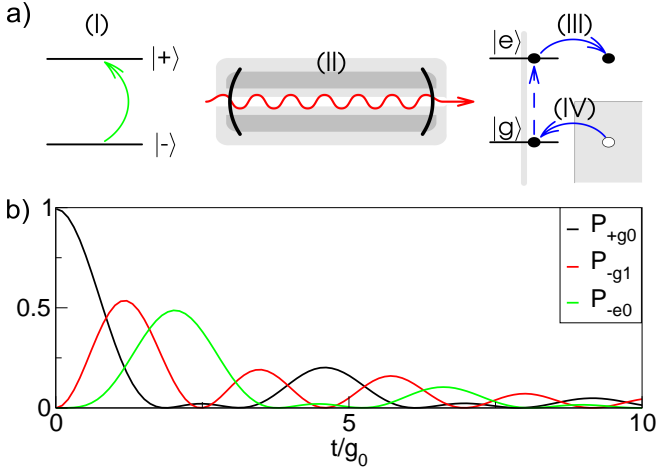} 
\caption{(color online) a) Complete excitation transfer cycle: (I) The
  qubit is excited by a fast pulse. (II) The qubit decays by emitting
  a photon into the cavity. (III) The dot electron absorbs the cavity
  photon and tunnels out to the lead. (IV) A lead electron tunnels
  into the ground state of the dot, leaving an electron-hole pair
  excitation in the lead. b) Decay of vacuum Rabi oscillations for
  $\Gamma=g_0$. Probabilities as a function of time for the states
  $|\! + \!g0\rangle, |\! -\! g1\rangle, |\! - \! e0\rangle$ shown.}
\label{pump}
\end{figure}

\textit{Single excitation dynamics --} To provide a basic picture of
the dynamics we first consider a single excitation of the isolated
DQD-cavity-qubit system. By applying an appropriate sequence of fast
pulses, on a nanosecond time scale, the qubit is excited, see
e.g. \cite{hofheinz09,Reed12}. This can be described by the operator
\(\hat R = \left( \hat \sigma_+ + \hat \sigma_- \right)/2 =\hat
\sigma_x \). Starting in the system ground state, exciting the qubit
gives $|\!- \! g0\rangle \rightarrow |\!+ \! g0\rangle$. Since $|\!+
\! g0\rangle$ is no eigenstate of the TC-Hamiltonian, following the
excitation at $t=0$ the system undergoes a free, coherent time
evolution described by the wavefunction
\begin{align}
\label{wf}
&|\Psi(t)\rangle=(1/\sqrt{2})\left[\cos^2\left(g_0t/\sqrt{2}\right) \vert \! + \! g0\rangle \right. \\ 
 &+ \left. \sin^2\left(g_0 t/\sqrt{2}\right)\vert \!- \! e0\rangle +(i/\sqrt{2})\sin\left(\sqrt{2}g_0t\right)\vert \!- \!  g1\rangle \right]. \nonumber 
\end{align}
This shows that the system performs vacuum Rabi oscillations, i.e. the
excitation oscillates between the qubit and the DQD, via the cavity,
with a frequency $\sqrt{2}g_0$.

Turning to the transport setup in Fig. \ref{fig_system}, with the DQD
coupled to a lead electrode, the result in Eq. (\ref{wf}) is modified.
The lead, at thermal equilibrium with a temperature \(T\) and chemical
potential \(\mu\), is described with the Hamiltonian \( \hat
H_L=\sum_k \epsilon_k \hat c^\dag_k \hat c_k\), where \(\hat
c^\dag_k\) denotes the creation operator for an electron at energy
\(\epsilon_k\). We will take the detuning between levels in the dots,
forming the DQD, to be zero. For this case the effective lead-DQD
tunneling amplitudes are equal for $|e\rangle$ and $|g\rangle$. The
tunnel Hamiltonian is then given by \(\hat H_T
=\sum_{k,\alpha=e,g}t_{k}\hat{c}^{\dag}_{k}\hat{d}_{\alpha}+H.c.\).

We consider the sequential tunneling regime \(\Gamma \ll \hbar
\omega\), with $\Gamma=2\pi \sum_k|t_k|^2\delta(\epsilon_k \pm \hbar
\omega_0/2)$ the DQD-tunneling rate. Applying the Born-Markov
approximation a quantum master equation (QME) \(d \hat{\rho}/dt =
\mathcal L \hat \rho \) is derived for the time evolution of the
reduced qubit-DQD-cavity system density operator $\hat{\rho}$. By
taking \(\mu=0\) and \(\hbar\omega_0/2 \gg k_B T \) the occupation of
the lead at the energies at which electrons tunnel into or out of the
DQD does not depend on the temperature. The Liouvillian $\mathcal{L}$
then becomes
\begin{equation}
\mathcal{L}\hat{\rho}=-\frac{i}{\hbar}[\hat{H}_{S},\hat{\rho}]+\frac{\Gamma}{2}\left[\mathcal{D}(\hat{\rho},\hat{d}^\dag_{e})+\mathcal{D}(\hat{\rho},\hat{d}_{g})\right],
\label{Liouvillian}
\end{equation}
with $\mathcal{D}(\hat{\rho},\hat{\gamma})=2\hat{\gamma}^{\dag}\hat{\rho}\hat{\gamma}-\hat{\gamma}\hat{\gamma}^{\dag}\hat{\rho}-\hat{\rho}\hat{\gamma}\hat{\gamma}^{\dag}$  for any operator $\hat \gamma$. 

The system dynamics after an excitation at $t=0$ is illustrated in
Fig. \ref{pump}. The vacuum Rabi oscillations decay, with the excited
state probabilities $\propto e^{-\Gamma t/4}$ for $\Gamma \ll
g_0$. Hence, the system excitation is eventually transformed into an
electron-hole pair excitation in the lead as $|\! - \! e0\rangle
\rightarrow |\! - \! 00\rangle \rightarrow |\! - \! g0\rangle$,
i.e. the DQD electron, in $|e\rangle$, tunnels out into the lead
followed by a back tunneling into $|g\rangle$.

\textit{Pump-probe scheme --} We consider an experimentally relevant
pump-probe scheme, with a periodic pumping, or excitation, of the
qubit. Importantly, the pump period $\tau$ constitutes a versatile
tool for investigating the different time scales of the dynamics. We
focus on periodic steady-state operation, where the state of the
system is entirely determined by the time $t$ passed since the last
pumping, i.e. \(\hat{\rho}(t + m \tau) = \hat{\rho}(t)\), with
$m=0,1,2..$ . The density operator is then given by
\begin{equation}
\hat{\rho}(t) = \E^{\mathcal{L}t}\mathcal{R}\E^{\mathcal{L}(\tau-t)}\hat{\rho}(t)\,,\quad 0 <t<\tau,
\label{eq_rhoss}
\end{equation}
where $\mathcal{R}\hat{\gamma}=\hat{R}\hat{\gamma}\hat{R}^{\dag}$.

The probe consists of a non-invasive charge detector (not shown in
Fig. \ref{fig_system}), monitoring the individual dot-lead tunneling
events in time \cite{Gustavsson06,Flindt09}. Since tunneling into (out
of) the DQD creates a hole (an electron) excitation in the lead, we
here consider the statistics of electron and hole transfers into the
lead per pumping cycle. The full counting statistics for a time
periodic system can be obtained along the lines of
\cite{Pistolesi-2004}. We introduce an electron (hole) counting field
\(\chi_e\) (\(\chi_h\)) into the Liouvillian of
Eq.~\eqref{Liouvillian}, so that \(\mathcal L \rightarrow \mathcal L
(\chi_e, \chi_h)\). The cumulant generating function (CGF) over \(N
\geq 1\) periods is
\begin{equation}
\mathcal S _N(\chi_e,\chi_h)=\ln[\langle[\mathcal R \E^{\mathcal L (\chi_e,\chi_h)\tau}]^N\rangle_0]\,,
\label{CGF}
\end{equation}
where
\(\langle\hat{\gamma}\rangle_{0}=\tr[\hat{\gamma}\hat{\rho}(0)]\). From
the CGF one obtains the mean and variance of the number of electrons
(holes) emitted per period as $\langle n_{e
  (h)}\rangle_{N}=(1/N)\partial S_{N}/\partial
(i\chi_{e(h)})|_{\chi_{e},\chi_{h}=0}$ and $\langle \delta n^{2}_{e
  (h)}\rangle_{N}=(1/N)\partial^{2}S_{N}/\partial
(i\chi_{e(h)})^{2}|_{\chi_{e},\chi_{h}=0}$. Moreover, the probability
of emitting $n_{e}$ electrons and $n_{h}$ holes per period, during $N$
periods, is given by \(P_N(n_{e},n_{h})=\int_{-\pi}^{\pi}d\chi_e
d\chi_h \E^{-iN(n_{e}\chi_e+n_{h}\chi_h)+S_{N}(\chi_{e},
  \chi_{h})}/(2\pi)^2 \).

\begin{figure}[t!]
\includegraphics[width=1.0\linewidth]{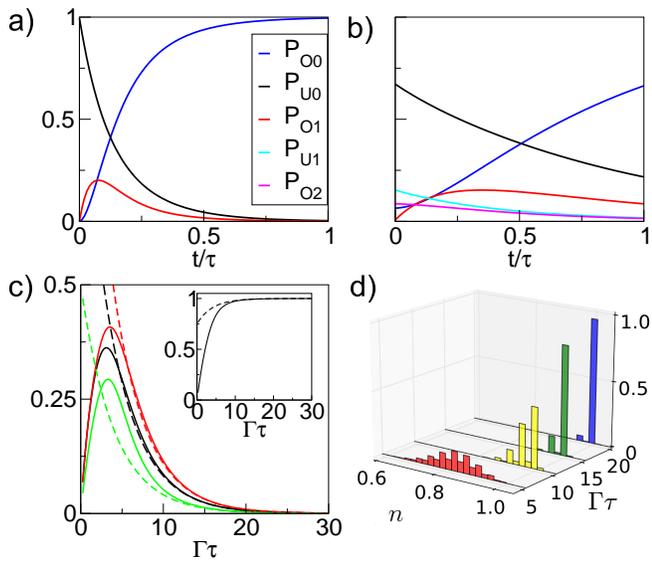}\label{results1a}
\caption{(color online) a), b) Time-dependent probabilities of the
  TC-eigenstates (see text) for complete, a), and incomplete, b),
  transfer cycles. Legend in b) same as in a). c) Electron (green
  lines) and hole (red lines) noise as a function of $\Gamma \tau$,
  for $N=1$. For $N\gg 1$ (black lines) electron and hole noise
  coincide. Inset: Average number of excitations $\langle n \rangle$
  as a function of $\Gamma \tau$. In both main panel and inset, the
  numerical (solid lines) and analytical, large $\Gamma \tau$ (dashed
  lines) results, are shown. d) Full probability distribution $P_N(n)$
  (for $N=50$) for different $\Gamma \tau$.}
\label{results1}
\end{figure}

\textit{Strong coupling regime.--} To illustrate in a compelling way
the proposed pump-probe scheme, we consider the strong coupling regime
\(g_0\gg \Gamma\) and a period $\tau \gg 1/g_0$
(Fig.~\ref{results1}). In this regime vacuum Rabi oscillations are
fast and can be averaged out in the density matrix (secular
approximation). The QME then reduces to an ordinary master equation in
the TC-basis, with the dynamics determined by $\Gamma$ and $\tau$. For
long periods \(\tau\gg1/\Gamma\) the system will describe a complete
cycle of transferring one energy quantum from the qubit, via a cavity
photon, to an electron-hole pair excitation in the lead (see
Fig.~\ref{pump}a). This picture is substantiated by the
electron-photon dynamics, illustrated by the corresponding time
dependent probabilities $P_{\sigma q}$ for the (set of)
TC-eigenstates, shown in fig.~\ref{pump}b. Here $\sigma$ denotes an
empty ($U$) or occupied ($O$) dot and $q$ the number of
excitations. We stress that in this long period regime the system acts
as an ideal electron-hole pair pump, or equivalently, a pump of energy
quanta $\hbar \omega$.

Decreasing the period $\tau$ there are corrections to the
complete transfer cycle. This is clear from the plot of the TC
probabilities in Fig.~\ref{pump}b: higher energy states with $q=2$,
including two cavity photons, acquire a finite population. Importantly,
the effect of finite $\Gamma \tau$ is clearly manifested in the full
counting statistics. A careful inspection of Eq. (\ref{eq_rhoss})
shows that the corrections scale as
\(x=\E^{-\Gamma\tau/4}\), much smaller than unity for large
$\Gamma\tau$. Solving the CGF to leading order in $x$ we get
\color{black}
\begin{equation}
\frac{S_{N}(\chi_{e},\chi_{h})}{N}=\I \left (\chi_{e}+\chi_{h}\right)+\frac{x(\E^{-2i(\chi_{e}+\chi_{h})}-1)}{8}+\frac{x\delta
S}{N}\,. \label{anaCGF}
\end{equation} 
The first term describes the complete cycle, while the second term
describes the first order corrections. The third term $\delta
S=(\cos(\chi_{h})-4)/3+[5\E^{-i(\chi_{e}+\chi_{h})}+3\E^{i(\chi_{e}+\chi_{h})}-(\E^{-2i(\chi_{e}+\chi_{h})}-1)]/8$
gives the short time behavior, going to zero for $N \gg 1$. Focusing
on the $N \gg 1$ limit, the electron and hole transfer statistics
become identical, with the number of transferred excitations
$n=n_e=n_h$, and we can write the full distribution (first order in
$x$)
\begin{equation}
P_N(n)=(1-x/8)\delta_{n1}+(x/8)\delta_{Nn,N-2}.
\label{anaProbs}
\end{equation}
From Eq. (\ref{anaProbs}) as well as from the plot of the numerically
evaluated $P_N(n)$, in Fig \ref{results1}, it is clear that the effect
of $x$ is i) a non-zero probability for $N-2$ excitations to be
transferred, and consequently, ii) a reduced probability for the ideal
cycle, transferring $N$ excitations. This can be explained as follows:
for non-zero $x$ there is a finite probability that an excitation is
pumped out of the system during a qubit ramp. The $P_N(n)$ to leading
order in $x$ describes one such event during \(N\) cycles. When an
excitation is pumped out of the system it does not generate an
electron-hole pair. Moreover, during the subsequent period the system
will remain in the ground state, where no electron-hole emission can
occur, thus resulting in a total loss of two periods for missed
excitations. We note that a similar mechanism was discussed in
\cite{Albert}.

For arbitrary $\Gamma \tau$, the numerically calculated probability
distribution $P_N(n)$ (for $N \gg 1$) as well as the two lowest
cumulants for the electron and hole transfer statistics are shown in
Fig.~\ref{results1}. The $P(n)$ shows that the ``missed two
excitations'' effect, for $\Gamma \tau \sim 1$, turns into a ``missed
several pairs of excitations'' effect, with a finite probability for
missing an even number of excitations $2,4,6, ...$. The probability to
miss an odd number of excitations is smaller, albeit increasing for
decreasing $\Gamma \tau$.

The average number of emitted electrons and holes are equal and
independent of $N$, $\langle n_e \rangle_N=\langle n_h
\rangle_N=\langle n \rangle$. In Fig. \ref{results1} $\langle n
\rangle$ is plotted together with the analytical solution for large
$\Gamma \tau$, $\langle n \rangle=1-\E^{-\Gamma\tau/4}/4$, following
from Eq. (\ref{anaCGF}). For \(\Gamma\tau\gtrsim 10\) the two results
coincide, while for smaller $\Gamma\tau$ the analytical solution
overestimates the number of emitted excitations.

In contrast to the average number, the electron and hole noise differ
for $N \sim 1$ while for $N \gg 1$ we find $\langle\delta
n^{2}_{e}\rangle_{N}=\langle\delta n^{2}_{h}\rangle_{N}\equiv \langle
\delta n^2\rangle$. As shown in Fig. \ref{results1}, for $\Gamma \tau
\gg 1$, both the electron and hole noise, for any $N$, are well
captured by the large $\Gamma \tau$ approximation, $\langle\delta
n^{2}_{e}\rangle_{N}=[(N+1)/2N]\E^{-\Gamma\tau/4}$ and $\langle\delta
n^{2}_{h}\rangle_{N}=[(3N+5)/6N]\E^{-\Gamma\tau/4}$, obtained from
Eq. (\ref{anaCGF}). The noise is thus exponentially suppressed when
the number of excitations per period (exponentially) approaches unity.
We point out that the difference between $\langle\delta
n^{2}_{e}\rangle_{N}$ and $\langle\delta n^{2}_{h}\rangle_{N}$ for $N
\sim 1$ is a consequence of electron and hole emission occurring at
different times during the pumping period. Hence, this electron-hole
asymmetry is altered if the measurement period is shifted with respect
to the pumping period. During the rest of the paper we consider the
$N\gg 1$ limit, when this asymmetry can be neglected. \color{black}

\begin{figure}[!t]
\includegraphics[width=0.49\linewidth]{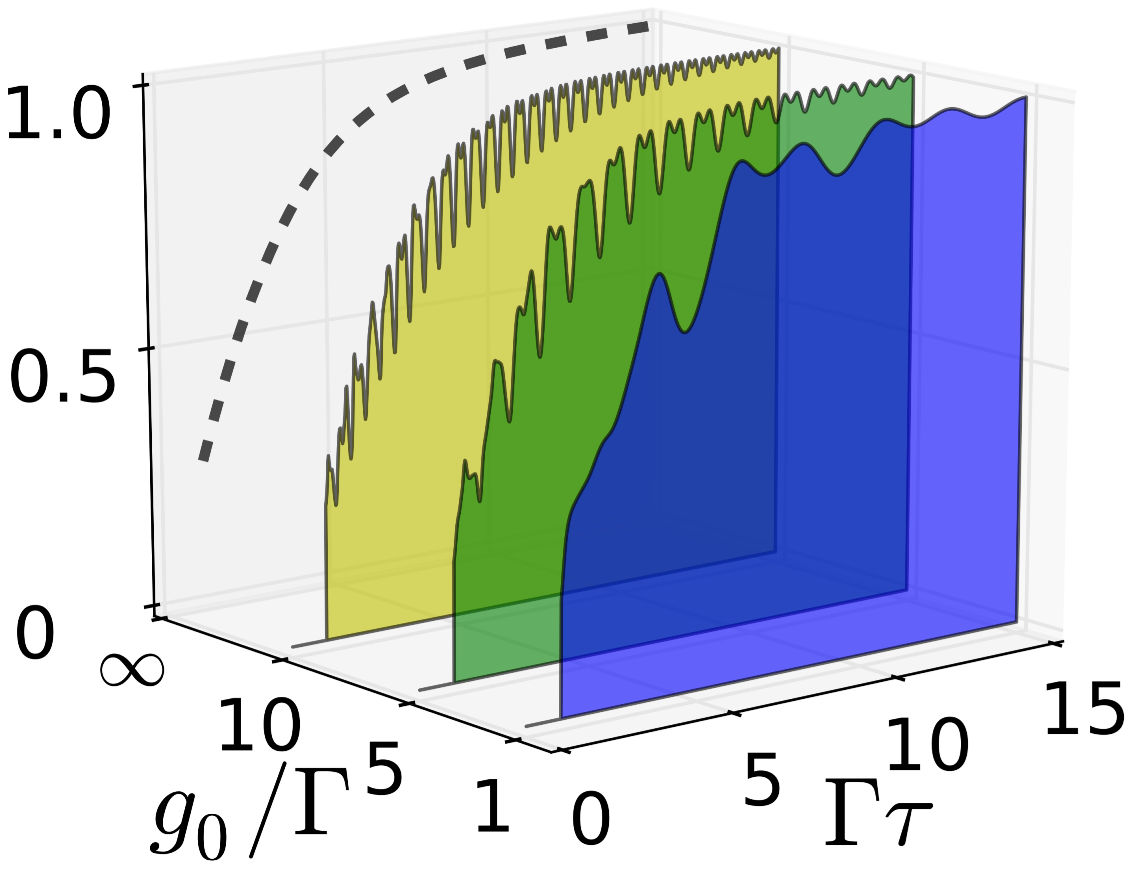}
\includegraphics[width=0.49\linewidth]{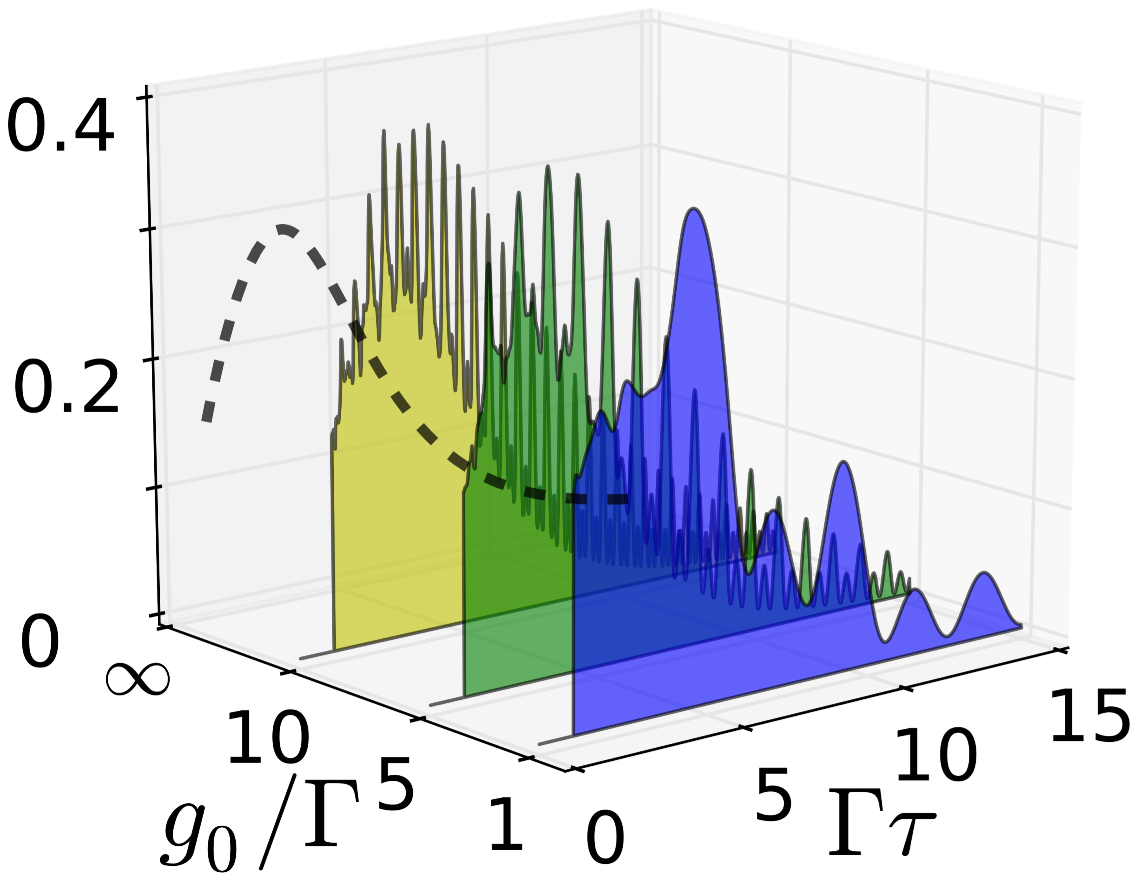}
\caption{\label{results2} (color online) Manifestation of vacuum Rabi oscillations. The average $\langle n \rangle$ (left panel) and fluctuations $\langle \delta n^2 \rangle$ (right panel) of number of transferred electron hole pairs per pumping period as a function of $\Gamma \tau$ for different $g_0/\Gamma$. Dashed line (in both panels) is strong coupling result ($g_0 \gg \Gamma$).}
\end{figure}

\textit{Vacuum Rabi oscillations --} Outside the strong coupling
regime, i.e. for $\Gamma\sim g_0$, the coherent electron-photon
dynamics becomes accessible. Focusing on period times $\tau$
comparable to $1/\Gamma,1/g_0$ signatures of vacuum Rabi oscillations
appear in the charge transfer statistics. This is clearly illustrated
in Fig. \ref{results2}, showing the average, $\langle n \rangle$, and
fluctuations, $\langle \delta n^2 \rangle$ of the number of
transferred electron-hole pairs per period, as a function of
$\Gamma\tau$. For both $\langle n \rangle$ and $\langle \delta n^2
\rangle$, moving out of the regime $g_0 \gg \Gamma$ the vacuum Rabi
oscillations are manifested as oscillations superimposed on the strong
coupling result.

The connection between the vacuum Rabi oscillations and the
superimposed oscillations in Fig. \ref{results2} can be illustrated by
considering two opposite cases of pumping periods: When $\tau \approx
2\pi n/[\sqrt{2}g_0]$ ($\tau \approx 2\pi (n+1/2)/[\sqrt{2}g_0]$) the
pumping occurs when the qubit has an increased probability to be
excited (in the ground state). As a consequence, the pumping will with
enhanced probability remove (add) an excitation, thereby decreasing
(increasing) the number of electron-hole pairs created.

\begin{figure}[h]
\includegraphics[trim=0cm 0cm 0cm 0cm, clip=true, width=0.98\linewidth]{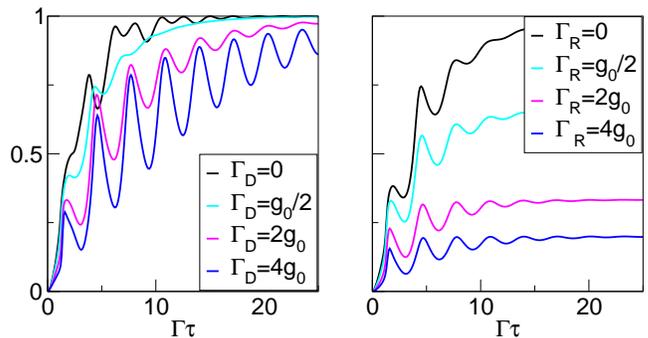}\label{results3b}
\caption{\label{results3} (color online) Effects of dephasing and relaxation. Average number of excitations $\langle n \rangle$ as a function of $\Gamma \tau$ for different dephasing rates $\Gamma_D$, with $\Gamma_R=0$ (left panel) and different relaxation rates $\Gamma_R$, with $\Gamma_D=g_0$ (right panel).}
\end{figure} 

\textit{DQD relaxation and dephasing --} In recent conductor-cavity
experiments
\cite{Frey112,Toida13,Petersson12,Viennot13,Deng13,Petta14} DQD
decoherence prevented reaching the strong coupling regime. This raises
the question how the signatures of coherent electron-photon
interaction in transport quantities are altered in the presence of
dephasing and relaxation. Dephasing and relaxation, with rates
$\Gamma_D$ and $\Gamma_R$ respectively, are accounted for by adding
the terms $\mathcal L_D [\hat \rho] = \frac{\Gamma_D}{2} \mathcal D
[\hat d^\dag_e \hat d_e - \hat d^\dag_g \hat d_g, \hat \rho]$ and
$\mathcal L_R [\hat \rho] = \frac{\Gamma_R}{2} \mathcal D [\hat
d^\dag_e \hat d_g, \hat \rho]$ to the Liouvillian in
Eq.~\eqref{eq_rhoss}.

The effect of dephasing is to suppress the coherence between the
excited, $\vert e \rangle$, and ground, $\vert g\rangle$, states of
the DQD. As illustrated in Fig. \ref{results3} this has several
consequences for the average number of transferred excitations
$\langle n \rangle$ (similarly for the fluctuations, not
shown). Increasing the dephasing from $\Gamma_D \ll g_0$ the amplitude
of the vacuum Rabi oscillations are first suppressed, reaching a
minimum around $\Gamma_D \sim g_0$. However, increasing the dephasing
further the oscillations are revived for $\Gamma_D > g_0$. In this
strongly dephased regime the system can be seen as a coherent
qubit-cavity subsystem, displaying vacuum Rabi oscillations with
frequency $2\sqrt{2}g_0$, coupled incoherently to the DQD. Moreover,
with increasing $\Gamma_D$ the typical excitation escape time
increases, i.e. the time it takes $\langle n \rangle$ to reach unity
gets longer.

To investigate the effect of relaxation we fix $\Gamma_D=g_{0}$ and
consider the dependence of $\langle n \rangle$ on $\Gamma_R$. The
results are presented in Fig. \ref{results3}. We see that the complete
cycle is no longer reached, i.e. $\langle n \rangle<1$ for $\Gamma
\tau \gg 1$. This can be explained by the fact that relaxation in the
DQD provides the system with a new mechanism of energy relaxation that
is independent of any pumping events. From this discussion it is clear
that dephasing and relaxation show up as distinct features in
transport properties when the period time $\tau$ is tuned.

\textit{Conclusions and estimates.--} In conclusion we have proposed
and analyzed a pump-probe scheme for transient electron-photon
dynamics in a hybrid DQD-cavity-qubit systems. We showed that key
features of the dynamics can be investigated via the short time charge
transfer statistics at the DQD. The dynamics was investigated in the
presence of DQD dephasing and relaxation, considering the
experimentally relevant situation with rates
$g_{0}\sim\Gamma_{D},\Gamma_{R}\sim 100$MHz. Note that accessing the
full transfer statistics on the time scale $1/g_0$ would require an
improved charge counting rate, presently below $1$MHz, and/or smaller
DQD dephasing and relaxation. The first two moments, current and
noise, could however be investigated by high-frequency measurements
\cite{Glattli}.

\textit{Acknowledgements--} We acknowledge support from the Swedish
VR and the nanometer structure consortium at Lund university. We also
thank Christian Flindt for providing constructive comments on the
manuscript.

\bibliography{bibHybrid}

\end{document}